\documentclass{article}

\usepackage[final, nonatbib]{neurips_2020}
\usepackage[utf8]{inputenc} 
\usepackage[T1]{fontenc}    
\usepackage{hyperref}       
\usepackage{url}            
\usepackage{booktabs}       
\usepackage{amsfonts}       
\usepackage{nicefrac}       
\usepackage{microtype}      
\usepackage{graphicx}
\usepackage{float}
\usepackage{xcolor}
\usepackage[backend=biber,sorting=none]{biblatex}
\title{Accurate and Scalable Matching of Translators to Displaced Persons for Overcoming Language Barriers}

\author{%
  Divyansh Agarwal\thanks{Delta Analytics; equal contribution.} \\
  \And
  Yuta Baba$^*$ \\
  \And
  Pratik Sachdeva$^{*,}$\thanks{University of California, Berkeley} \\
  \And
  Tanya Tandon$^{*}$ \\
  \And
  Thomas Vetterli$^{*}$\thanks{Correspondence to: \texttt{thomas.vetterli@gmail.com}} \\
  \And 
  Aziz Alghunaim\thanks{Tarjimly} \\
}

\begin{document}

\maketitle

\begin{abstract}
Residents of developing countries are disproportionately susceptible to displacement as a result of humanitarian crises. During such crises, language barriers impede aid workers in providing services to those displaced. To build resilience, such services must be flexible and robust to a host of possible languages. \textit{Tarjimly} aims to overcome the barriers by providing a platform capable of matching bilingual volunteers to displaced persons or aid workers in need of translating. However, Tarjimly's large pool of translators comes with the challenge of selecting the right translator per request. In this paper, we describe a machine learning system that matches translator requests to volunteers at scale. We demonstrate that a simple logistic regression, operating on easily computable features, can accurately predict and rank translator response. In deployment, this lightweight system matches 82\% of requests with a median response time of 59 seconds, allowing aid workers to accelerate their services supporting displaced persons.
\end{abstract}

\section*{Introduction}
In 2019, at least 79.5 million people were forcibly displaced out of their houses due to humanitarian crises, which have disproportionately impacted countries in the developing world \cite{un_report}. Building resilience during these crises requires multiple organizational layers of support for such displaced persons, often necessitating communication across languages. Since interpreters are a scarce and expensive resource, language barriers heavily impede the ability of aid workers to provide support for displaced people. \textit{Tarjimly} \cite{tarjimly} aims to lower these barriers by leveraging the reach and scaling power of a free mobile application to bridge the gap between the world's 3 billion bilingual individuals and displaced persons in need of language support. Specifically, Tarjimly provides a platform in which a volunteer translator can participate in a live chat session with a displaced person or aid worker, in which they can send texts, voice notes, documents, or start a live phone call and provide translation for the requester. Thus, Tarjimly helps humanitarian organizations build resilience by democratising language support when interpreters are not available.

Tarjimly has consolidated a group of roughly 20,000 translators worldwide, providing a critical mass of volunteers to serve the nearly 30,000 displaced persons and aid workers on its platform. However, ensuring that the numerous daily translator requests are quickly and accurately matched to translators presents a significant challenge. Given the large pool of volunteers, notifying each one per translation request is not feasible. Rather, targeting a subset of translators, ranked according to their probability of response, is a more effective and scalable approach. Thus, machine learning can serve as a possible tool to quickly match requests to translators at scale. In this work, we demonstrate that a machine learning system composed of a logistic regression accurately predicts translator responsiveness to requests, using easily computable features in real-time and at scale.

\section*{Methods}
\subsection*{Data Generation}
\begin{figure}[t]
    \centering
    \includegraphics[scale=0.8]{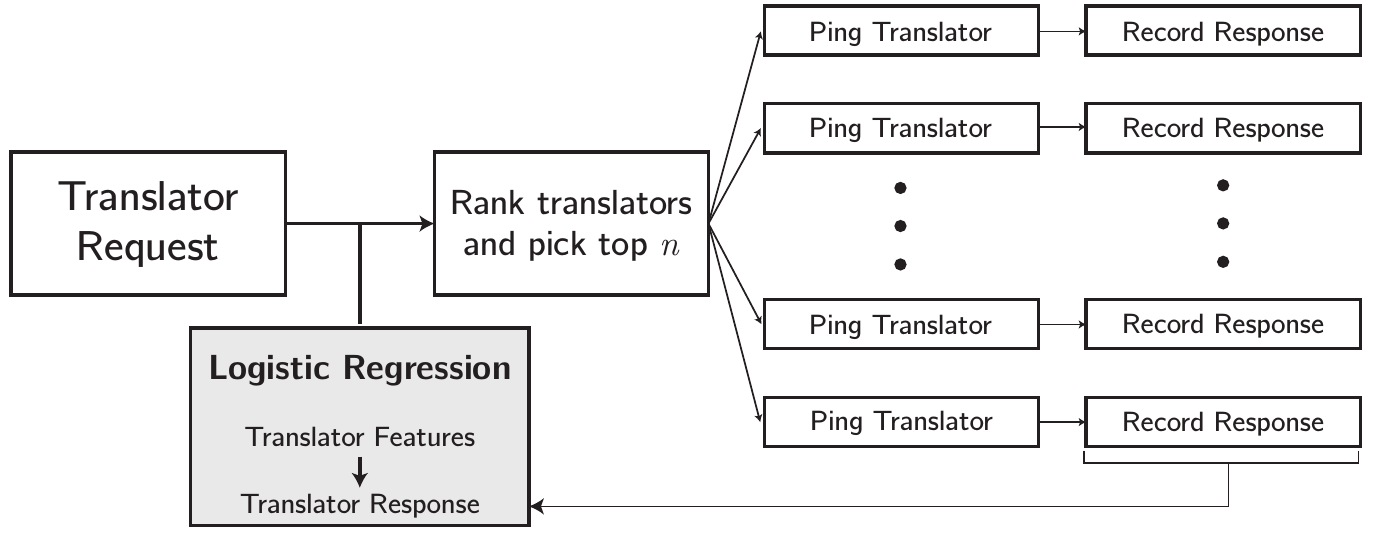}
    \vspace{-5pt}
    \caption{\textbf{Translator matching pipeline}. Each request for a translator is passed through a logistic regression classifier after a series of filtering steps. The top $n=30$ translators are ranked according to the classifier, and pinged. Translator responses (yes, no, and null response) are recorded for further training of the classifier. The first translator to respond ``yes'' is matched with the requester.}
    \label{fig:ping_pipeline}
    \vspace{-10pt}
\end{figure}

The translator matching pipeline relies on two components: a mobile application on which users sign up as translators or beneficiaries, as well as a Django backed micro-service that processes and matches the translator requests via trained machine learning models. These models were served and analyzed using the \texttt{mlflow} \cite{mlflow} platform. For every translation request, the system filters translators according to specific criteria, including language choice, timezone, and requester preferences (e.g., preference for a particular gender identity of occupation). After filtering, the machine learning algorithm ranks translators by probability of response to the request. The top $n$ (typically 30-60) translators are ``pinged'' via a notification on the app (Fig.~\ref{fig:ping_pipeline}). Translators can either accept a ping, decline, or do nothing. The translator that provides the first affirmative response to a ping is matched to the requester. All requests, pings, and translator responses are stored for use as training data in a binary or multi-class classification (e.g., positive, negative, or null response) problem.

\subsection*{Translator Response Classifier}
We modeled the matching task as a binary classification, with the positive class ($y=1$) corresponding to a response of ``yes'', and the negative class ($y=0$) corresponding to a response of ``no'', or lack of a response. During deployment, the classifier's predicted probabilities rank translators according to those most likely to respond. We use a logistic regression classifier \cite{cox1958regression} for its simplicity, ease of maintenance in production, and model interpretability.

The model operates on two broad categories of features: historical response rates and translator profile features. For ping $p$ and translator $t$, we calculated the ``overall response rate'' $r_p^{(t)}$ as $r_p^{(t)} = \sum_{i=1}^{p-1} \frac{y_i^{(t)} + 1}{p+2}$, where $y_i^{(t)}$ is translator $t$'s response to ping $i$. The quantity $r_p^{(t)}$ is $0.5$ for translators with no pings and asymptotes to 1 (0) for the perfectly responsive (unresponsive) translator. Furthermore, we included a ``periodic response rate,'' calculated similarly as $r_p^{(t)}$, but restricted to a translator's history in the hour of day for the current ping. Meanwhile, the translator profile features consist of self-reported translating experience (an ordinal variable quantifying ``formal translator experience'' and a binary feature denoting ``ability to translate documents''), a binary feature denoting whether the translator explicitly declared they were available at the time of ping (``availability''), and a binary feature indicating whether the translator has translated in multiple professional settings (e.g., medicine, media, education: ``multi-skill'').

\section*{Results}
\begin{figure}[t]
    \centering
    \includegraphics[scale=1.2]{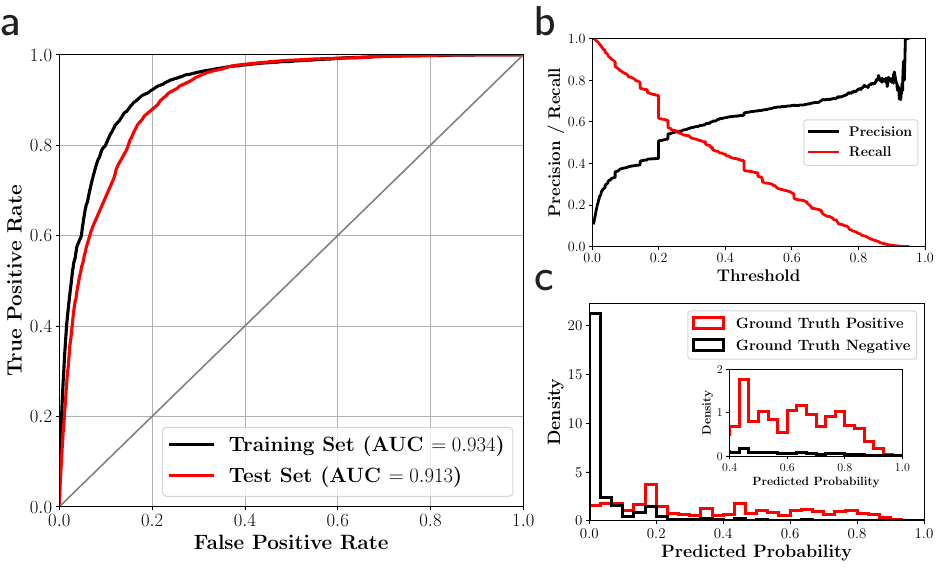}
    \caption{\textbf{Classification performance using response rate and translator profile features.} \textbf{a.} Receiver operator characteristic (ROC) curves for the training set (black) and test set (red). \textbf{b.} Precision (black) and recall (red) as a function of classification threshold, evaluated on the test set. \textbf{c.} The density distribution of the translator response probabilities calculated on test set samples. Distributions are divided by ground truth label with the positive (translator responds yes) as red and negative (translator responds no/doesn't respond) as black. Inset highlights the density distribution for likely positive responses (predicted probability of $0.4$ or higher).}
    \label{fig:results}
    \vspace{-10pt}
\end{figure}

We trained the logistic regression classifier on a dataset consisting of roughly 400,000 pings, with 20\% of the data held out as a test set. We used an $\ell_2$ penalty to regularize the classifier, and chose the regularization strength via temporal cross-validation \cite{hyndman2018forecasting, ramachandran2020} while optimizing for the area under the curve (AUC) on the validation data. The dataset was heavily imbalanced, with about 6\% of samples in the total dataset occurring in the positive class (translator responding ``yes''). However, we required no special imbalance learning techniques to achieve sufficient predictive quality. 

Our results for a classifier trained using both response rate and translator profile features are summarized in Figure \ref{fig:results}. Since the goal of the classifier in deployment is to rank translators, we assessed predictive performance by examining the receiver operator characteristic (ROC) curve (Fig.~\ref{fig:results}a). We found that the classifier excelled at identifying responsive translators, obtaining an AUC of $0.91$ on the test set, indicating superior predictive quality across thresholds. This translated to a raw accuracy of $91\%$ on the test set, using a standard threshold of $0.5$ (relative to $89\%$ achieved by a naive classifier universally predicting the negative class).

In \textit{Tarjimly's} use case, false negatives (predicting a translator will not respond, when they will) are more detrimental than false positives (predicting a translator will respond, when they will not), as the latter may spur engagement on future requests. Thus, we examined the precision-recall curves (Fig.~\ref{fig:results}b). At the standard threshold of $0.5$, the classifier achieved a precision of $0.65$ and recall of $0.34$, implying moderate control of false positives and false negatives. This was reflected in the clear separation of predicted probabilities between the ground truth positive and negative examples in the test set (Fig.~\ref{fig:results}c), particularly for the high thresholds likely to be used in deployment (Fig.~\ref{fig:results}c: inset). Since recall degrades in this regime, an adaptive threshold could be used to reduce false negatives. 

\begin{figure}[t]
    \centering
    \includegraphics[scale=0.36]{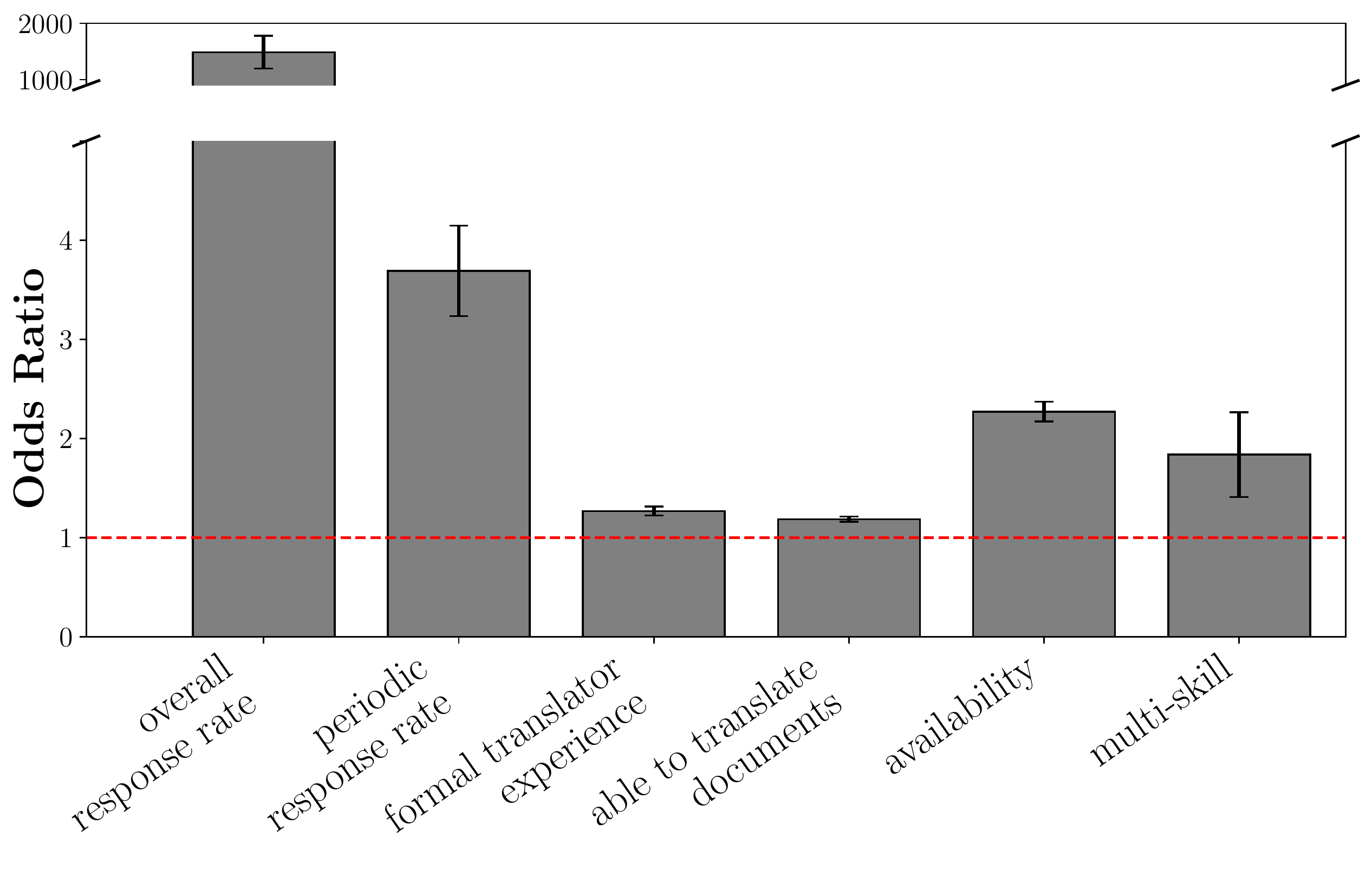}
    \vspace{-15pt}
    \caption{\textbf{Feature importance.} Odds ratios (exponential of model coefficients) per feature. Red dashed line denotes an odds ratio of 1, where the feature does not impact the calculated probability of translator response. Note that the $y$-axis is broken due to the odds ratio of the overall response rate. Error bars denote standard deviation across cross-validation folds.}
    \label{fig:odds_ratios}
    \vspace{-10pt}
\end{figure}

To assess feature importance, we examined the coefficients of the trained logistic classifier. Specifically, we calculated the odds ratio, or the exponential of the coefficient values. An odds ratio greater than 1 indicates that the presence of or increase in a feature corresponds to a higher probability of the translator responding ``yes'', according to the classifier. The odds ratios for the trained classifier are depicted in Figure~\ref{fig:odds_ratios}. We found that, by far, the most important features were the response rate features (overall and periodic). In particular, the overall response rate dominated the other features, with an odds ratio orders of magnitude larger than the other features. However, the classifier still utilized predictive signal across the feature set, with all odds ratios greater than 1 (positively predictive). These feature importances imply that historically responsive, highly skilled, and specially trained translators are the likeliest to respond affirmatively to a request, in line with our expectations.

Since the response rate features dominated during prediction, we evaluated how a classifier using only translator profile features would perform on the data. A translator profile classifier was moderately predictive, achieving an AUC of $0.63$ (ROC curves not shown) with qualitatively similar feature importances as in Figure~\ref{fig:odds_ratios}. Thus, the response rate features are instrumental in prediction, likely needing correction to ensure that responsive translators are not overused during requests.

To ensure that our algorithm performed consistently across different segments of our users, we analyzed the classification performance across language pairs. In an initial analysis, we found that classification performance was relatively stable across language pairs with greater than 1000 samples, which suggested that a language-independent classifier could be used as an initial model. We aim to further study the behavior for language pairs with fewer samples as they become populated.

\section*{Discussion}
In this work, we demonstrated that a simple machine learning algorithm can effectively match translators to displaced persons. Furthermore, our approach utilizes easily computable features, easing the scalability of the algorithm during deployment. However, successful deployment requires multiple goals beyond the predictive performance for which we optimized. In production, key metrics of interest include the match rate (fraction of requests that result in a match), match time (how long it takes to match), and the user rating (how well the translator served the needs of the request). While the former two metrics are correlated with the task we constructed, the latter is not incorporated in the model. Validating that our trained model achieves good performance on \textit{all} metrics during deployment is of paramount importance. A preliminary analysis of the models in production has shown that they achieve a match rate of 82\% with a median response time of 59 seconds. 

We employed a logistic regression for its simplicity, scalability, and ease of deployment. However, other commonly used machine learning algorithms, such as decision trees and random forests, may provide additional insight into which translator features predict responsiveness. In addition, imbalance learning techniques might further improve predictive accuracy \cite{he2009learning, chawla2002smote}.

Our approach heavily relies on response rate features, calculated from the historical activity of a translator (Fig.~\ref{fig:odds_ratios}). Despite their predictive capacity, these features can result in the over-utilization of active translators and under-utilization of newer or less active ones. Thus, traversing the exploration-exploitation problem in this context is of interest for future work \cite{katehakis1987multi}. To this end, we are currently developing an exploration-exploitation paradigm via the epsilon greedy algorithm \cite{banditsbook}. Under this approach, we randomly choose translators to ping with probability $\epsilon$, while matching according to our trained model with probability $1 - \epsilon$. While we chose this approach due to its simplicity, other approaches, such as Thompson scaling, are also of interest for future work \cite{thompson1933likelihood}.

\section*{Broader Impact}
Tarjimly's machine learning system provides translation services to the millions of displaced persons who have access to a mobile phone either personally or via an aid worker. The power of crowd-sourcing lies in its reach and flexibility, which are particularly useful for supporting existing systems that provide aid, and reaching individuals who may have fallen through the cracks of these systems.

An immediate challenge in our current approach lies in our heavy reliance on historical features, which can potentially lead to overuse of a small group of highly active translators. While this may result in a high match rate, it only serves to reduce the flexibility of the application, while making engagement of new translators more difficult. In the long term, this may lead to high turnover on the platform, ultimately decreasing the overall accuracy of the matching procedure. In an effort to mitigate these side effects, we are utilizing an epsilon greedy algorithm to encourage exploration of new or under-utilized translators. Furthermore, we are working to identify predictive signatures of translator re-engagement.

Our feature analysis suggests an archetype of a responsive translator as a highly trained or skilled individual with the flexibility to respond to translator requests. While socioeconomic status is not a feature used by our algorithm, this archetype implies that many translators heavily relied upon by our matching procedure may be of high socioeconomic status. Ultimately, we prioritized the needs of the displaced person or aid worker benefiting from the service provided by the volunteer translator. Thus, it is worth assessing the fairness of the matching algorithm insofar as it impacts the success of the translation services. If so, further analyses examining whether a metric of interest should be equalized across a protected class could be conducted.

An additional benefit of the platform lies in its ability to provide support to individuals that may speak a less common language or dialect. However, rarer language pairs may be less represented in the training data because there are few translators speaking those languages or requests for that pair.  Ensuring that the system can properly serve communities that speak rare languages is necessary to ensure that its impact is fair across users.

A deeper, more general issue lies in how \textit{Tarjimly} fits in the broader realm of providing services to displaced peoples. The usage of crowd-sourced translators poses the risk of transmitting information or translations without the proper nuance or accuracy. The consequences of incorrect information during such crises can be severe. Additionally, ensuring that translators are able to ethically translate sensitive materials without infringing on privacy is imperative to ensuring the safety of the users. Thus, \textit{Tarjimly} does not aim to serve as a replacement for on-the-ground interpreters, who can offer more nuanced and meaningful support to displaced persons, with proper oversight. Instead, it aims to fill the gaps when interpreters are not available, providing support to bolster resilience. Therefore, \textit{Tarjimly} aims to deepen its ties with humanitarian organizations in an effort to better situate itself in this broader ecosystem, providing translation services effectively and safely. 

\section*{Acknowledgments}
This work was completed as part of the 2020 Delta Analytics\footnote{Delta Analytics (http://www.deltanalytics.org) is a non profit whose aim is to pair San Francisco,CA professionals with local non-profits.} Data Fellowship. During the fellowship, DA, YB, PS, and TT served as Data Fellows, and TV served as project lead. We thank Atif Javed and the rest of Tarjimly for their feedback and support on methods development and integration within the Tarjimly platform. Additionally, we thank Terence Tam, Arman Madani, Ugaso Sheikh-Abdi and Saud Alsaif for laying the groundwork on the previous iteration of the models described in this paper. We thank the executive board and all members of Delta Analytics for their support during the Data Fellowship as well as their useful feedback on our approach. Lastly, we thank the anonymous reviewers for their substantive and thoughtful comments.
\clearpage
\printbibliography

\end{document}